\pdfoutput=1
\documentclass[a4paper,aps,prd,10pt,preprintnumbers,showpacs,onecolumn,superscriptaddress,nofootinbib,amsmath,amssymb,preprint]{revtex4-1}
\usepackage{graphicx}
\usepackage{cmap}
\usepackage[utf8]{inputenc}
\usepackage[T1]{fontenc}
\usepackage{soul}
\usepackage{amsmath}
\usepackage{hyperref}
\usepackage{epstopdf}
\begin{document}

\title{Long-lived quasinormal modes in the Euler-Heisenberg electrodynamics}

\author{B. C. Lütfüoğlu}
\email{bekir.lutfuoglu@uhk.cz}
\affiliation{Department of Physics, Faculty of Science, University of Hradec Kralove, \\
Rokitanskeho 62/26, Hradec Kralove, 500 03, Czech Republic.}

\begin{abstract}
Using the precise Leaver method and time-domain integration, we analyze the quasinormal modes and late-time behavior of massive neutral and charged scalar fields in the background of a charged, asymptotically flat black hole in the presence of Euler–Heisenberg nonlinear electrodynamics. We show that as the field mass increases, the damping rate decreases significantly, approaching arbitrarily long-lived states known as quasi-resonances. However, these modes cannot be identified in time-domain profiles due to the dominance of asymptotic tails in this regime, which decay slowly and exhibit oscillations with a power-law envelope. We observe that a larger field charge leads to a significantly higher quality factor, as it increases the oscillation frequency while reducing the damping rate.
\end{abstract}

\maketitle

\section{Introduction}

Despite the wide range of nonlinear electrodynamics (NED) models proposed in the literature, many suffer from either the lack of a consistent weak-field limit or being introduced merely as phenomenological constructions. In contrast, the Euler--Heisenberg electrodynamics \cite{Heisenberg:1936nmg} provides a physically well-founded framework, since it arises as a low-energy effective description of quantum electrodynamics (QED). One of its most striking predictions is the existence of light-by-light (photon--photon) scattering in vacuum, a process absent in classical Maxwell theory, whose theoretical basis was established in \cite{Karplus:1950zz} and later confirmed through detailed analyses and collider experiments \cite{dEnterria:2013zqi, TOTEM:2021zxa}. When coupled to Einstein gravity, this theory leads to the Einstein--Euler--Heisenberg (EEH) model, in which charged black holes deviate from the standard Reissner--Nordström geometry in the strong-field regime. Although the corrections are typically small, they can exert a profound impact on the dynamics of perturbations, as demonstrated in several studies of stability and classical properties \cite{Bolokhov:2024ixe, Nomura:2021efi, Breton:2021mju, Breton:2016mqh}. Whereas neutral fields are likely to be only moderately influenced by the nonlinear character of the underlying electrodynamics, charged fields are expected to be strongly affected.

Quasinormal modes (QNMs) govern the characteristic oscillations of black holes and uniquely encode information about their parameters and the underlying gravitational theory \cite{Konoplya:2011qq, Kokkotas:1999bd, Bolokhov:2025uxz}. In gravitational-wave astronomy, the detection of QNMs provides a direct test of the no-hair theorem and possible deviations from general relativity in the strong-field regime. Beyond astrophysical applications, QNMs also play a key role in gauge–gravity duality, where they determine relaxation timescales in strongly coupled quantum field theories.

So far, most investigations of quasinormal spectra in the EEH background have been restricted to \emph{massless} fields, primarily examining fundamental modes while neglecting the massive case. Yet, the introduction of mass is known to enrich the spectrum dramatically, giving rise to long-lived modes and even quasi-resonances in various contexts. To date, a systematic study of massive perturbations in EEH black holes---especially of their higher overtones and the quasi-resonant regime---has been essentially absent.  In what follows, we provide a motivation for focusing on massive fields in black hole spacetimes, highlighting the physical scenarios where their QNMs play a central role.

The investigation of QNMs of massive fields has attracted continuous attention in recent decades due to the rich phenomenology that emerges once a nonvanishing mass term is introduced \cite{Konoplya:2004wg, Ohashi:2004wr, Ponglertsakul:2020ufm, Konoplya:2006br, Gonzalez:2022upu, Burikham:2017gdm, Malik:2024voy, Konoplya:2018qov, Deng:2025uvp, Konoplya:2017tvu, Becar:2025niq, Chen:2025sbz, Zhidenko:2006rs, Konoplya:2005hr, Zhang:2018jgj, Aragon:2020teq}. A notable aspect is that the effective mass often arises naturally in a variety of settings. For example, in braneworld scenarios, the influence of the higher-dimensional bulk can induce an effective mass term in the perturbation equations \cite{Seahra:2004fg}. Moreover, the possibility of a massive graviton, whether within massive gravity theories or as an emergent effective degree of freedom, has strong implications for the interpretation of very long-wavelength gravitational waves. Such waves are now under close scrutiny by Pulsar Timing Array collaborations, including NANOGrav, which have recently reported evidence for a correlated stochastic background \cite{Konoplya:2023fmh, NANOGrav:2023hvm}.

Another key motivation comes from the distinctive dynamical properties of massive field perturbations. For particular ranges of the field mass, QNMs can become arbitrarily long-lived, leading to so-called quasi-resonances \cite{Ohashi:2004wr, Konoplya:2004wg}. This phenomenon is quite general, encompassing perturbations of different spin fields \cite{Konoplya:2005hr, Fernandes:2021qvr, Konoplya:2017tvu, Percival:2020skc}, a variety of black hole geometries \cite{Konoplya:2006br, Konoplya:2013rxa, Zhidenko:2006rs, Zinhailo:2018ska, Konoplya:2019hml, Churilova:2020bql, Bolokhov:2023bwm}, and even exotic compact objects such as wormholes \cite{Churilova:2019qph}. The late-time behavior of massive perturbations is also qualitatively different from the massless case: instead of the usual monotonic power-law tails, one finds oscillatory decays, a feature that has been extensively analyzed in \cite{Jing:2004zb, Koyama:2001qw, Moderski:2001tk, Rogatko:2007zz, Koyama:2001ee, Koyama:2000hj, Gibbons:2008gg, Gibbons:2008rs}. Furthermore, even fields that are intrinsically massless may acquire an effective mass when propagating near black holes immersed in magnetic fields \cite{Konoplya:2007yy, Konoplya:2008hj, Wu:2015fwa, Kokkotas:2010zd}, thereby broadening the range of physical scenarios where massive-field QNMs are relevant. Yet, the presence of quasi-resonances is not universal: examples are known where massive fields do not support arbitrarily long-lived modes \cite{Zinhailo:2024jzt, Konoplya:2005hr}, underscoring the importance of case-by-case analysis. These diverse motivations highlight the central role of massive-field perturbations in probing both gravitational theory and astrophysical phenomena.

When discussing electrically charged black holes, the most interesting phenomenon occurs when taking into account the interaction between the black hole charge and the charge of a field under consideration. Consequently, extensive literature exists on charged-field QNMs and late-time tails. A comprehensive analysis of massive charged scalar perturbations in Kerr–Newman spacetimes is given by Konoplya and Zhidenko, who derive analytic tail expressions and study mode behavior near extremality \cite{Konoplya:2013rxa}. In the non-rotating case, Richartz extends the study to large-charge regimes by deriving QNM formulas for both scalar and Dirac fields \cite{Richartz:2014jla}, and González et al. uncover anomalous decay behaviors in de Sitter settings \cite{Gonzalez:2022upu}. Importantly, the universal oscillatory late-time decay law \(t^{-5/6}\sin(\mu t)\), first confirmed numerically by Burko and Khanna (2004), applies across different spins and geometries \cite{Burko:2004jn}, including vector fields \cite{Konoplya:2006gq}.  

Motivated by these considerations, in this work, we study the QNMs and time-domain evolution of a charged scalar field in the background of a charged black hole within the framework of Euler--Heisenberg NED.

The paper is organized as follows. In Section~\ref{sec2}, we introduce the background geometry of the charged black hole in Euler--Heisenberg NED and formulate the perturbation equations for a massive charged scalar field. Section~\ref{sec3} presents the numerical methods employed--the Frobenius (Leaver) method, the time-domain integration scheme, and the WKB approach with Padé approximants--as well as the numerical results, including the quasinormal spectra, late-time tails, and the occurrence of quasi-resonances. Finally, in the Conclusions, we summarize our findings and outline possible future extensions of this work.

\section{Black Hole Geometry and Scalar Perturbation Equations} \label{sec2}

In order to investigate the dynamics of massive scalar fields, we consider a class of charged black holes supported by NED described by the Euler--Heisenberg effective action. Unlike purely phenomenological models of NED, the Euler--Heisenberg Lagrangian arises naturally from QED as the leading correction due to vacuum polarization in strong electromagnetic fields. When coupled to general relativity, this framework gives rise to static, spherically symmetric black holes that deviate from the standard Reissner--Nordström solution in the high-curvature regime near the event horizon, while preserving asymptotic flatness.

The dynamics follow from the action \cite{Magos:2020ykt}
\begin{equation}
S = \frac{1}{4\pi} \int_{M^{4}} d^{4}x \sqrt{-g}\left[ \frac{1}{4} R   +F - \frac{4\alpha^2}{45m_e^4}\,F^{2} - \frac{7\alpha^2}{45m_e^4}\,G^{2} \right],
\end{equation}
where $R$ is the Ricci scalar, $g$ denotes the determinant of the metric, and the electromagnetic invariants are
$F = \tfrac{1}{4}F_{\mu\nu}F^{\mu\nu}$ and
$G = \tfrac{1}{4}F_{\mu\nu}\,{}^{\ast}F^{\mu\nu}$.
The parameter $\alpha$ governs the strength of the Euler--Heisenberg correction.

For a purely electric configuration, the resulting line element can be written as
\begin{equation}
ds^{2} = -f(r)\,dt^{2} + \frac{dr^{2}}{f(r)} + r^{2} d\Omega^{2},
\end{equation}
with the lapse function
\begin{equation}
f(r) = 1 - \frac{2M}{r} + \frac{Q^{2}}{r^{2}} - \frac{aQ^{4}}{r^{6}},
\label{eq:EHmetric}
\end{equation}
and the electric field
\begin{equation}
A_{\mu}dx^{\mu}=-\left(\frac{Q}{r}-2a\frac{Q^3}{r^5}\right)dt,
\label{eq:EHelectric}
\end{equation}
where $M$ and $Q$ denote the ADM mass and the electric charge of the black hole, and $a$ is a constant proportional to $\alpha$, given by
\begin{equation}
    a =\frac{32\alpha^2}{9m_e^4}.
\end{equation}
The last term in~\eqref{eq:EHmetric} encodes the NED correction, which becomes relevant only in the strong-field regime.

\begin{figure}
\resizebox{\linewidth}{!}{\includegraphics{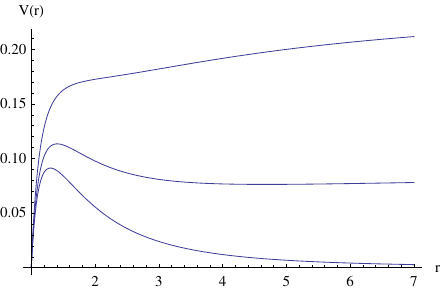}~\includegraphics{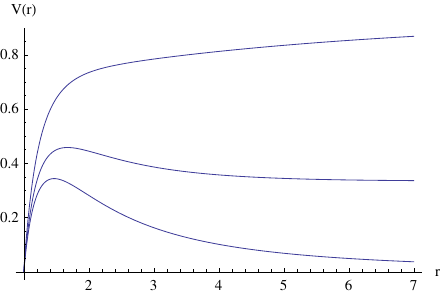}}
\caption{Effective potential for scalar perturbations. Left panel: $\ell=0$, $r_{h}=1$, $Q=0.5$, $a=1$, $q=0$, $\mu=0$ (bottom), $\mu=0.3$ and $\mu=0.5$ (top). Right panel: $\ell=1$,  $r_{h}=1$, $Q=0.5$, $a=0.5$, $q=0$, $\mu=0$ (bottom), $\mu=0.6$, $\mu=1$ (top).}\label{fig:Potentials}
\end{figure}

The propagation of a massive charged scalar field in this background is governed by the Klein--Gordon equation
\begin{equation}
\Phi_{;\mu}^{\ \ ;\mu} - i q A_\mu g^{\mu\nu} \left( 2 \Phi_{;\nu} - i q A_\nu \Phi \right) - i q A_{\mu\ ;\nu} g^{\mu\nu} \Phi = \mu^2 \Phi,
\end{equation}
with $q$ and $\mu$ are, respectively, the scalar field charge and mass. Decomposing the field into spherical harmonics and isolating the radial part,
\begin{equation}
\Phi(t,r,\theta,\varphi) = \frac{\Psi(r)}{r} Y_{\ell m}(\theta,\varphi)e^{-i\omega t},
\end{equation}
one obtains a Schrödinger-like wave equation for $\Psi(r)$ in the tortoise coordinate $r^{*}$, defined through $dr^{*}/dr = 1/f(r)$:
\begin{equation}
\frac{d^{2}\Psi}{dr^{* 2}} + \left[ \left(\omega-\frac{qQ}{r}+2a\frac{qQ^3}{r^5}\right)^{2} - V(r) \right]\Psi = 0.
\end{equation}
Here, the effective potential for the massive neutral scalar field takes the form
\begin{equation}
V(r) = f(r)\left( \frac{\ell(\ell+1)}{r^{2}} + \frac{f'(r)}{r} + \mu^{2} \right),
\label{eq:Vscalar}
\end{equation}
where $\ell$ denotes the multipole number. The term proportional to $\mu^{2}$ is responsible for qualitatively new features of massive-field perturbations, including the appearance of quasi-resonant modes and altered late-time tails, phenomena that will be analyzed in subsequent sections. The effective potentials for neutral fields are shown in Fig.~\ref{fig:Potentials}. They are positive definite, which guarantees the stability of perturbations, that is, the decaying character of all QNMs. 

For nonzero field charge $q$, the effective potential becomes a complex-valued function due to the coupling term, and it also acquires an implicit dependence on the quasinormal frequency $\omega$. This feature is generic for charged perturbations in the presence of an electromagnetic background and implies that the standard interpretation of the potential as a fixed scattering barrier is no longer strictly applicable.

\section{Quasinormal modes and late-time evolution of perturbations} \label{sec3}

Without loss of generality we consider  $\mathrm{Re}(\omega)> 0$. However, in the spectrum we have both positive and negative values of $\mathrm{Re}(\omega)$, which can be obtained by replacing $q \mapsto-q $, $\omega \mapsto - \omega^{*}$ \cite{Patrick:2018orp}.  This replacement does not change the imaginary part of $\omega$, so that the dominant frequencies for positive and negative $q$ have the same decay rate.   

In the asymptotically flat case, the QNM boundary conditions require
\begin{equation}
\Psi(r) \propto
\begin{cases}
e^{- i \tilde{\omega} r^{*}}, & r \to r_{+} \qquad (\text{purely ingoing at the horizon}),\\[4pt] e^{+ i \Omega r^{*}}, & r \to \infty \qquad (\text{purely outgoing at infinity}),
\end{cases}
\label{eq:BCs}
\end{equation}
where
\begin{equation}
\Omega = \sqrt{\omega^{2} - \mu^{2}},
\qquad \tilde{\omega} = \omega - \frac{q Q}{r_{+}}+2a\frac{q Q^3}{r_{+}^5}.
\end{equation}
The square root is chosen such that $\mathrm{Re}(\Omega)> 0$,  so that (\ref{eq:BCs})  ensures a purely outgoing wave at infinity, since we consider $\mathrm{Re}(\omega)> 0$ .

Following Leaver's method \cite{Leaver:1985ax}, one factors out the singular behavior at the boundaries by writing
\begin{equation}
\Psi(r) = \left( \frac{r - r_{+}}{r - r_{-}} \right)^{- i \tilde{\omega} / (4 \pi T_{H})}
e^{i \Omega r} (r-r_{-})^{i \sigma} \, y(r),
\label{eq:factorization}
\end{equation}
where
\begin{equation}
\sigma = \left( \Omega + \frac{\mu^{2}}{2\Omega} \right) (r_{+} + r_{-}),
\end{equation}
and $T_{H}$ is the Hawking temperature,
\begin{equation}
T_{H} = \frac{r_{+} - r_{-}}{4 \pi r_{+}^{2}},
\end{equation}
while $y(r)$ is regular at both $r = r_{+}$ and $r \to \infty$. The function $y(r)$ is expanded as a Frobenius series around the event horizon:
\begin{equation}
y(r) = \sum_{k=0}^{\infty} a_{k} \left( \frac{r - r_{+}}{r - r_{-}} \right)^{k}.
\label{eq:Frobenius_series}
\end{equation}

This series converges for $r_{+} \leq r < \infty$ and automatically satisfies the boundary condition at the horizon. The QNM condition is enforced by requiring convergence also at spatial infinity.

Substitution of \eqref{eq:Frobenius_series} into the radial equation yields usually a three-term recurrence relation for the coefficients:
\begin{equation}
\alpha_{n} a_{n+1} + \beta_{n} a_{n} + \gamma_{n} a_{n-1} = 0,
\quad n \geq 0,
\quad \gamma_{0} = 0,
\label{eq:recurrence_relation}
\end{equation}
where $\alpha_{n}$, $\beta_{n}$, and $\gamma_{n}$ depend on $\omega$ and the black hole parameters.

The convergence of the series at spatial infinity is ensured if the recurrence coefficients satisfy the continued-fraction equation:
\begin{equation}
\beta_{n} - \frac{\alpha_{n-1} \gamma_{n}}{\beta_{n-1} -}
\frac{\alpha_{n-2} \gamma_{n-1}}{\beta_{n-2} - \cdots}
= \frac{\alpha_{n} \gamma_{n+1}}{\beta_{n+1} -}
\frac{\alpha_{n+1} \gamma_{n+2}}{\beta_{n+2} - \cdots}.
\label{eq:continued_fraction}
\end{equation}
In practice, this is solved numerically by minimizing the absolute difference between the left- and right-hand sides. The infinite fraction is truncated at a sufficiently large index $n_{\mathrm{max}}$.  In some cases, a higher-order recurrence relation arises, which is treated in the same way.

The Nollert improvement \cite{Nollert:1993zz} accelerates the convergence of the continued fraction, especially for modes with large imaginary parts, by using the asymptotic expansion of the ratio
\begin{equation}
R_{n} \equiv \frac{a_{n+1}}{a_{n}} = C_{0} + \frac{C_{1}}{\sqrt{n}} + \frac{C_{2}}{n} + \cdots,
\label{eq:Nollert_asymptotics}
\end{equation}
to approximate the tail of the fraction. This reduces the required truncation order and improves numerical stability in slowly convergent cases.

The Frobenius (Leaver) method, supplemented by the Nollert improvement, yields numerically precise QNM spectra for massive and charged scalar fields in charged black hole backgrounds. It is applicable to both fundamental modes and higher overtones, and works equally well for massless and massive cases, provided that the proper asymptotic form \eqref{eq:BCs} is used (see, e.g., \cite{Konoplya:2004uk, Dias:2022oqm, Stuchlik:2025mjj, Xiong:2023usm, Kanti:2006ua, Zinhailo:2024kbq, Konoplya:2007zx, Stuchlik:2025ezz}).

To study the dynamical response of a charged scalar field in the background of a static, asymptotically flat charged black hole, we employ the time-domain integration method. In the tortoise coordinate $r^{*}$, the perturbation equation can be written as
\begin{equation}
\label{td_eq}
\left(\frac{\partial^{2}}{\partial {r^{*}}^{2}}
 - \frac{\partial^{2}}{\partial t^{2}}
 + 2 i\,\Phi(r^{*})\,\frac{\partial}{\partial t}
 - V(r^{*})\right)\Psi(t,r^{*})=0,
%\tag{18}
\end{equation}
where the effective potential $V(r)$ and the electromagnetic coupling term $\tilde{\Phi}(r)$ are given by
\begin{equation}
\label{V_phi_defs}
V(r)= f(r)\!\left[\frac{\ell(\ell+1)}{r^{2}}+\frac{f'(r)}{r}+\mu^{2}\right]-\tilde{\Phi}(r)^{2},
\qquad
\tilde{\Phi}(r)=-\frac{qQ}{r}+2a\frac{qQ^3}{r^5},
\end{equation}
with $f(r)$ being the metric function, $\mu$ the field mass, $q$ the field charge, and $Q$ the black hole charge. The tortoise coordinate $r^{*}$ is defined via $dr^{*}/dr = f(r)^{-1}$, ensuring that spatial infinity and the event horizon are mapped to $r^{*}\to+\infty$ and $r^{*}\to-\infty$, respectively.

We discretize \eqref{td_eq} on a grid $(t_i, r^{*}_j)$ with steps $\Delta t$ and $\Delta r^{*}$ inside an equilateral triangle. The finite-difference scheme for the charged scalar field reads \cite{Zhu:2014sya}
\begin{equation}
\label{FD_scheme}
\begin{aligned}
\Psi_{j,i+1}
&= \frac{(1+i\Phi_j \Delta t)\,\Psi_{j-1,i}+(2-\Delta t^{2} V_j)\,\Psi_{j,i}}
        {1-i\Phi_j \Delta t} \\
&\quad +\frac{\Delta t^{2}}{\Delta {r^{*}}^{2}}\,
\frac{\Psi_{j+1,i}-2\Psi_{j,i}+\Psi_{j-1,i}}
     {1-i\Phi_j \Delta t},
\quad
r^{*}_j=r^{*}_0+j\Delta r^{*},\ \ t_i=t_0+i\Delta t,
\end{aligned}
%\tag{20}
\end{equation}
where $\Phi_j \equiv \tilde{\Phi}(r_j)$ and $V_j \equiv V(r_j)$.

Once the time-domain profile $\Psi(t,r^{*}_{\text{obs}})$ is obtained at a fixed observation point $r^{*}_{\text{obs}}$, the dominant quasinormal frequencies can be extracted by fitting the signal to a sum of damped exponentials using the Prony method:
\begin{equation}
\label{Prony}
\Psi(t)\simeq \sum_{k=1}^{p} C_k\,e^{-i\omega_k t}, 
%\tag{21}
\end{equation}
where $C_k$ are complex amplitudes and $\omega_k$ are the complex quasinormal frequencies, with the real parts corresponding to the oscillation frequencies and imaginary parts determining the damping rates. We place the initial Gaussian wave packet near the peak of the effective potential in order to avoid a prolonged initial outburst. We then verify that shifting the fitting window to later stages of the quasinormal ringing does not change the extracted frequencies within the required accuracy. However, when the field mass becomes sufficiently large, the late-time tail begins very early, and in this regime we are unable to extract the quasinormal frequencies with adequate precision. We are also limited by the low accuracy in time domain for $\ell=0$ perturbations, because the ringdown phase consists only from a few oscillations in that case. Since we impose only the initial conditions and no external source, we automatically have a purely outgoing wave in the asymptotic region and the infalling wave to the black-hole event horizon.

The time-domain integration, together with the Prony method, is frequently used for the analysis of spectra and stability of black holes and other compact objects showing excellent agreement with more accurate methods \cite{Aneesh:2018hlp, Bronnikov:2021liv, Dubinsky:2024hmn, Dubinsky:2024jqi, Bronnikov:2019sbx, Abdalla:2012si, Varghese:2011ku, Churilova:2021tgn, Cuyubamba:2016cug, Dubinsky:2025azv, Skvortsova:2024wly, Skvortsova:2024atk, Skvortsova:2024eqi, Skvortsova:2023zmj, Konoplya:2013sba, Bolokhov:2023dxq, Lutfuoglu:2025ohb, Lutfuoglu:2025hwh}.

In our computations, the WKB method \cite{Schutz:1985km, Iyer:1986np, Konoplya:2003ii} with Padé approximants \cite{Matyjasek:2017psv} is employed solely to provide an initial guess for the quasinormal frequencies, which is then refined using the Frobenius (Leaver) method. This semi-analytic approach approximates the solution of the wave equation by matching asymptotic expansions near the potential peak, with Padé resummation improving the convergence for low-lying modes. This method has been widely used for finding QNMs and grey-body factors in numerous publications (see, for example \cite{Matyjasek:2021xfg, Kodama:2009bf, Malik:2023bxc, Malik:2024nhy, Konoplya:2005sy, Konoplya:2001ji, Bolokhov:2023ruj, Paul:2023eep, Lutfuoglu:2025hjy, Konoplya:2021ube, Dubinsky:2025fwv, Bolokhov:2025lnt}). While the method offers reasonable accuracy for moderate and high multipole numbers, especially in the eikonal regime, we refer the reader to \cite{Konoplya:2019hlu, Matyjasek:2017psv} for detailed descriptions and its range of applicability. When the field mass becomes so large that the effective potential no longer develops a peak, we iteratively use previously computed quasinormal frequencies as initial guesses for the modes at larger values of $\mu$.

The configuration under consideration involves several parameters, including the charges $Q$ and $q$, the black-hole mass $M$, the multipole number $\ell$, the field mass $\mu$, and the nonlinearity parameter $a$. Providing exhaustive numerical data covering the full range of all these parameters would require substantial space. Therefore, for illustrative purposes, we select representative values of the black-hole parameters that adequately demonstrate the characteristic shifts in the quasinormal frequencies. Our main emphasis is placed on the dependence on the field mass $\mu$, since this parameter gives rise to the most interesting phenomena, such as quasi-resonances and the distinctive late-time behavior.

\begin{figure}[h!]
\resizebox{\linewidth}{!}{\includegraphics{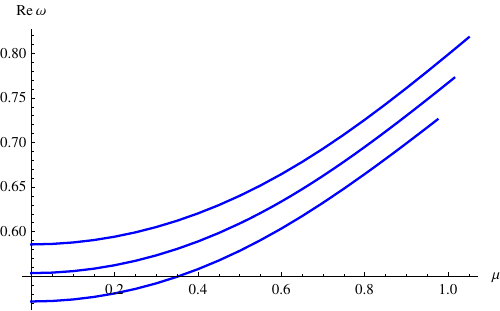}~~~\includegraphics{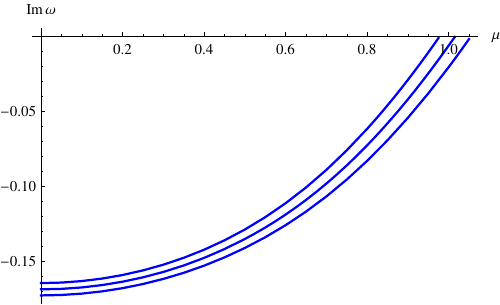}}
\caption{Imaginary part of $\omega$ (right panel), $n=0$ for $\ell=1$ scalar perturbations: $M=1$, $Q=0.5$, $a=0.5$, $q=-0.1$, $n=1$ (top), $q=0$, $n=0$ (middle) $q=0.1$, $n=0$ (bottom).  Real part of $\omega$ for the same values of the parameters (left panel); $q=-0.1$ (bottom), $q=0$ (middle), $q=0.1$ (top). Notice that in the spectrum there are both positive and negative values of $\mathrm{Re}(\omega)$, which can be obtained by replacing $q \mapsto-q $, $\omega \mapsto - \omega^{*}$. }\label{fig:L1n0}
\end{figure}

\begin{figure}
\resizebox{\linewidth}{!}{\includegraphics{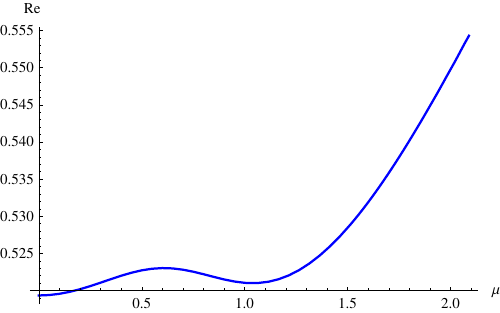}~~~\includegraphics{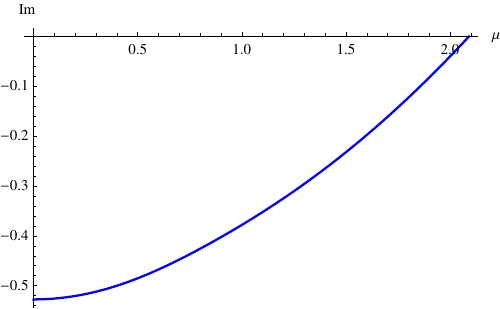}}
\caption{Real (left panel) and imaginary (right panel) parts of $\omega$ for the first overtone $n=1$ with $\ell=1$ scalar perturbations: $M=1$, $Q=0.5$, $a=0.5$, and $q=0$.} \label{fig:L1n1}
\end{figure}

\begin{figure}
\resizebox{\linewidth}{!}{\includegraphics{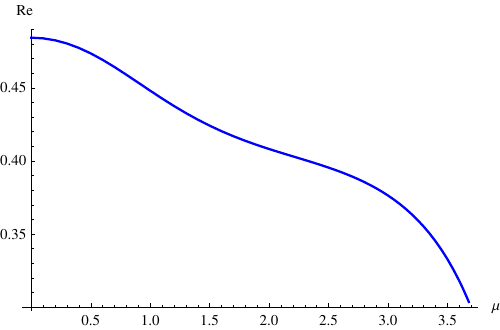}~~~\includegraphics{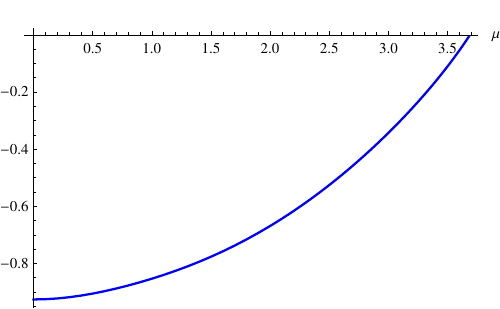}}
\caption{Real (left panel) and imaginary (right panel) parts of $\omega$ for the second overtone $n=2$ with $\ell=1$ scalar perturbations: $M=1$, $Q=0.5$, $a=0.5$, and $q=0$.} \label{fig:L1n2}
\end{figure}

For both neutral and charged fields, we observe the phenomenon of quasi-resonances: as $\mu$ increases, the damping rate $\mathrm{Im}(\omega)$ approaches zero at a certain critical mass, yielding arbitrarily long-lived modes, as shown in Figs.~\ref{fig:L1n0}-\ref{fig:L1n2}.


\newpage
When the fundamental mode reaches this quasi-resonant state, the first overtone becomes the new fundamental mode. The real oscillation frequency $\mathrm{Re}(\omega)$ of this new fundamental mode displays a pronounced minimum near the critical value of $\mu$. However, we do not find an analogous behavior at larger $\mu$, when the second overtone replaces the first, suggesting that the observed minimum is a numerical coincidence rather than a general pattern.

The field charge $q$ also has a clear impact on the spectrum. Increasing $q$ leads to a higher damping rate and a lower real oscillation frequency, as shown in Fig.~\ref{fig:L1n0} for positive $q$. 

\begin{figure}
\centerline{
\resizebox{0.33\linewidth}{!}{\includegraphics{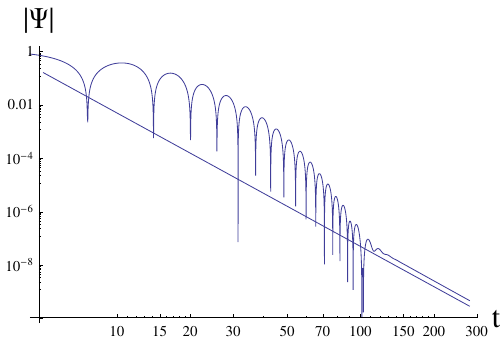}}
\resizebox{0.33\linewidth}{!}{\includegraphics{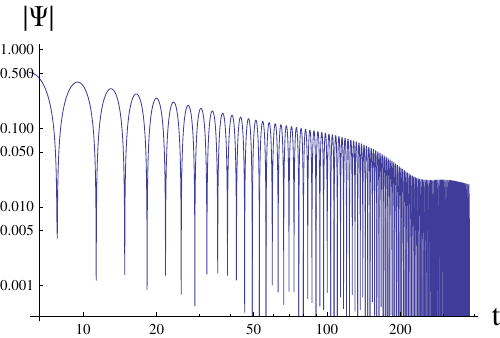}}
\resizebox{0.33\linewidth}{!}{\includegraphics{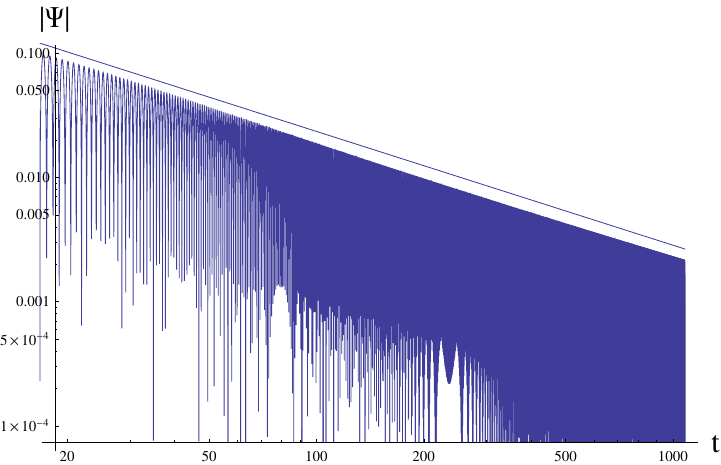}}}
\caption{The logarithmic time-domain profile for $\ell=1$ scalar perturbations: $r_{h}=1$, $Q=0.5$, $a=0.5$, $q=0$, $\mu=0$, (left), $\mu=1$, $r_{h}=1$ (middle) and $M=1$, $\mu=5$ (right). For the massless case on the left plot, the asymptotic late-time tail is $\sim t^{-5}$. The Prony fit gives $\omega = 0.553432 - 0.16884 i$, while the precise Leaver method gives $\omega = 0.553431 - 0.168843 i$. For the massive case $\mu=1$ and $\mu=5$, the long-lived QNM $\omega =0.766690 - 0.006224 i$ cannot be extracted because of the early dominance of the intermediate late-time tails. Here, we used the units $r_{h}=1$ for the first two plots to allow for the straightforward comparison with QNMs obtained by the Leaver method. The last plot is obtained in units $M=1$, which is convenient for comparison with the known asymptotic tails. }\label{fig:TD1}
\end{figure}

\begin{figure}
\centerline{
\resizebox{0.5 \linewidth}{!}{\includegraphics{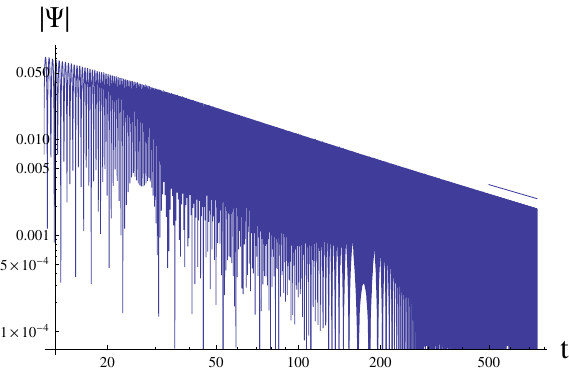}}
\resizebox{0.5 \linewidth}{!}{\includegraphics{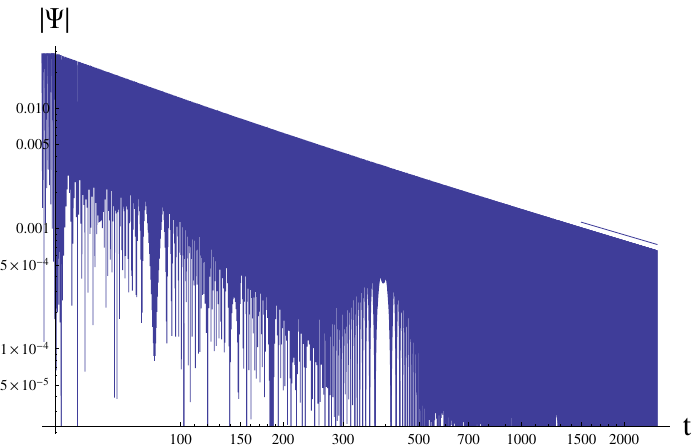}}}
\caption{The logarithmic time-domain profile for $\ell=0$, $a=5$ (left) and $\ell=1$, $a=10$ (right) scalar perturbations: $M=1$, $Q=0.5$, $q=0$, $\mu=10$. The asymptotic decay law is $\sim t^{-5/6}$. }\label{fig:TD2}
\end{figure}

\begin{figure}
\centerline{
\resizebox{0.5 \linewidth}{!}{\includegraphics{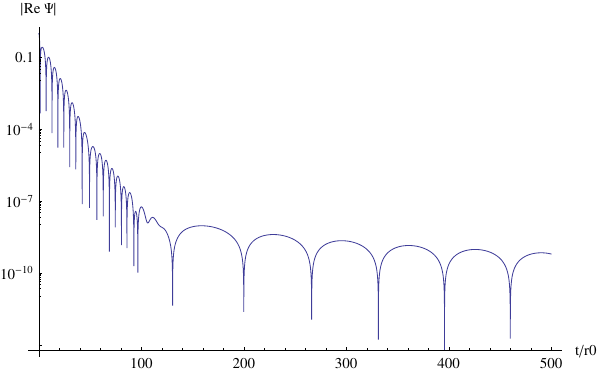}}
%\resizebox{0.5 \linewidth}{!}{\includegraphics{TDEHMassiveL1mu10e0.pdf}}
}
\caption{The semi-logarithmic time-domain profile for $\ell=1$ scalar perturbations: $r_{h}=1$, $a=0.5$, $Q=0.5$, $q=0.1$, $\mu=0.05$. The Prony method gives $\omega = 0.586188 - 0.172546 i$, while the Frobenius method gives $\omega =0.585609 - 0.172905 i$.}\label{fig:TD3}
\end{figure}

For the massless case, the late-time behavior of the neutral scalar field is governed by the well-known Price law \cite{Price:1971fb}, exhibiting a power-law decay (see the left plot in Fig.~\ref{fig:TD1}):
\begin{equation}
\Psi(t) \propto t^{-(2\ell+3)},
\end{equation}
with no oscillatory component.

In the limit of the Schwarzschild or Reissner-Nordstr\"om black hole, when the field is massive, the asymptotic regime changes qualitatively: the decay becomes oscillatory with an inverse power-law envelope \cite{Koyama:2001ee},
\begin{equation}
\Psi(t) \propto t^{-5/6} \, \sin(\mu t + \delta),
\end{equation}
where $\mu$ is the field mass and $\delta$ is a constant phase shift. For the NED under consideration, we see that the late-time tails are almost the same; still, the best fit in units $M=1$, for example, for $\mu=5$, $\ell =1$ in Fig.~\ref{fig:TD1} is
\begin{equation}
\Psi(t) \propto t^{-11/12} \, \sin(\mu t + \delta).
\end{equation}
This may indicate either that the asymptotic tail contains a subdominant correction arising from the nonlinearity parameter $a$, being a perturbative parameter, is expected to be relatively small, or that the true asymptotic regime has not yet been reached.  To exclude the former possibility, we chose an artificially large coupling, $a=5$, and a large field mass, $\mu=10$, at $\ell =0$, so that the asymptotic tails would appear earlier. On the other hand, if the correction due to $a$ were non-zero, it should be visible under these conditions. As shown in Fig.~\ref{fig:TD2}, the asymptotic tail $\sim t^{-5/6}$ dominates for $\ell=0$ and $\ell=1$ respectively. Notice that at intermediate time, the best fit for these plots is again $\sim  t^{-11/12}$, that is, the asymptotic regime dominates at sufficiently late-times.

For $q \neq 0$, the late-time signal is affected by the complex nature of the potential, leading to a modulation of the oscillatory tail. The imaginary component of the potential introduces an additional phase shift and a mild damping in the envelope of the oscillations. Although the overall power-law decay rate remains comparable to the neutral case, the phase evolution becomes charge-dependent, a feature that could, in principle, be used to distinguish charged from neutral fields in numerical simulations or observational modeling. From Fig.~\ref{fig:TD3}, we see that the time-domain integration and the Frobenius method are in very good concordance, whenever the mass $\mu$ is small enough and the asymptotic tails begin at sufficiently late-times, allowing for a long period of quasinormal ringing.

\section{Conclusions} \label{sec:concl}

We have computed the quasinormal spectra and late-time behavior of massive neutral and charged scalar fields propagating in the background of a charged, asymptotically flat black hole in the Einstein–Euler–Heisenberg theory. Using the Frobenius method and time-domain integration, we have shown that increasing the field mass leads to a strong suppression of the damping rate, giving rise to arbitrarily long-lived states — quasi-resonances. However, in the time-domain profiles, such modes are hidden by the dominance of slowly decaying oscillatory tails with a power-law envelope. For charged fields, the interplay between the black hole charge and the field charge significantly affects both the quasinormal spectrum and the late-time decay rate. Our results extend previous analyses of the Reissner–Nordström case to a physically well-motivated nonlinear electrodynamics background, providing a first step toward the study of charged massive fields in this framework.

\acknowledgments
The author is grateful to Excellence Project PrF UHK 2205/2025-2026 for the financial support.

\bibliography{bibliography}
\end{document}